\documentclass[a4paper, conference]{IEEEtran}

\usepackage{url}

\usepackage{units}

\usepackage{amsmath}
\usepackage{graphicx}

\usepackage{etoolbox}
\makeatletter
\patchcmd{\@makecaption}
  {\scshape}
  {}
  {}
  {}
\makeatletter
\patchcmd{\@makecaption}
  {\\}
  {.\ }
  {}
  {}
\makeatother
\begin{document}

\title{Secure Parallel Processing of Big Data Using Order-Preserving Encryption on Google BigQuery}

\author{\IEEEauthorblockN{Timo Schindler}
\IEEEauthorblockA{
Faculty of Computer Science\\ and Mathematics\\
Laboratory for Information Security\\
OTH Regensburg\\
timo.schindler@oth-regensburg.de}
\and
\IEEEauthorblockN{Christoph Skornia}
\IEEEauthorblockA{Faculty of Computer Science\\ and Mathematics\\
Laboratory for Information Security\\
OTH Regensburg\\
christoph.skornia@oth-regensburg.de}}

\maketitle
\thispagestyle{plain}
\pagestyle{plain}

\begin{abstract}
With the increase of centralization of resources in IT-infrastructure
and the growing amount of cloud services, database management systems (DBMS) 
will be more and more outsourced to Infrastructure-as-a-Service (IaaS) 
providers. The outsourcing of entire databases, or the computation power 
for processing Big Data to an external provider also means that the 
provider has full access to the information contained in the database.
In this article we propose a feasible solution with Order-Preserving 
Encryption (OPE) and further, state of the art, encryption methods to sort and process 
Big Data on external resources without exposing the unencrypted data to the
IaaS provider. We also introduce a proof-of-concept client for Google BigQuery as 
example IaaS Provider.
\end{abstract}

\section{Introduction to Order Preserving Encryption}
Cloud Computing has reached to be one of the cornerstones in IT-Infrastructure during 
the last years. Especially the outsourcing of databases is one main service and has 
been proposed in several publications. 
\cite{Agrawal:2012:ELD:2180777.2180778, amazon, Brantner:2008:BDS:1376616.1376645, relational-cloud-mit, Das:2013:EES:2445583.2445588, google-cloud-db, microsoft-azure} 
The idea to use complex networking and computing infrastructure as a 
service is reasonable but has its limitations. 
\cite{994695, Hacigumus:2002:ESO:564691.564717} An external platform of data storage 
has to be treated as untrusted. Encryption is a powerful technology for 
protecting the confidentiality of the data stored but needs to be decrypted for 
processing. One approach is to use encryption which allows operations on encrypted 
data. Fully Homomorphic Encryption (FHE) and Order-Preserving Symmetric Encryption 
(OPE) are relevant approaches to solve this dilemma with encryption algorithms.
As OPE maintains the order of the encrypted data obtained, data can be compared on 
the encrypted system and are thus sorted. Since the comparison of data is already 
sufficient to run a significant amount of common operations on the remote 
database system, this method fullfills two important prerequisites for the 
outsourcing of DBMS: 
Data can not only be securely stored, but also processed on a remote system.
Because of this quality, OPE is primarily used in databases for processing SQL 
queries over encrypted data \cite{compare, Agrawal:2004:OPE:1007568.1007632, 4221716, Hacigumus:2002:ESO:564691.564717, secure-relation-database, 2009-chaotic, 6253544, CPE:CPE2992, Popa:2011:CPC:2043556.2043566, xiao-extending}. 
The concept of OPE is subject of research since a number of years and secure 
algorithms were found. \cite{compare, Popa:2011:CPC:2043556.2043566, Mavroforakis:2015:MOE:2723372.2749455} The challenge for researchers is to develop a 
feasible solution for relevant use cases of secure data processing. 
Some of the most recent aspects are considered and evaluated in this work.

In the context of the outsourcing process of computing power and services to 
external, untrusted systems, Fully Homomorphic Encryption is an 
alternative to OPE. In contrast to OPE, FHE offers the 
advantage that the homomorphic property of the encrypted data is retained. 
Thus, an OPE scheme only guarantees that comparison's plaintext space ($x>y$) 
has the same result as those obtained in the encrypted space ($Enc \left(x
\right)>Enc\left(y\right)$). In FHE, computation of more sequential 
operations on encrypted data are possible. Recent work on FHE has shown it is, in 
principal, possible the perform arbitrary computations over encrypted data 
\cite{Gentry:2009:FHE:1536414.1536440}, the performance overheads are prohibitively 
high, on the order of $10^9$ times \cite{compare, Gentry2012}. Both procedures have 
their specific use case and can be used together. In this work we focus on OPE as we 
are mainly interested in the option to sort data on the untrusted IaaS-Platform.

\begin{figure*}
\noindent \begin{centering}
\includegraphics[width=0.95\textwidth]{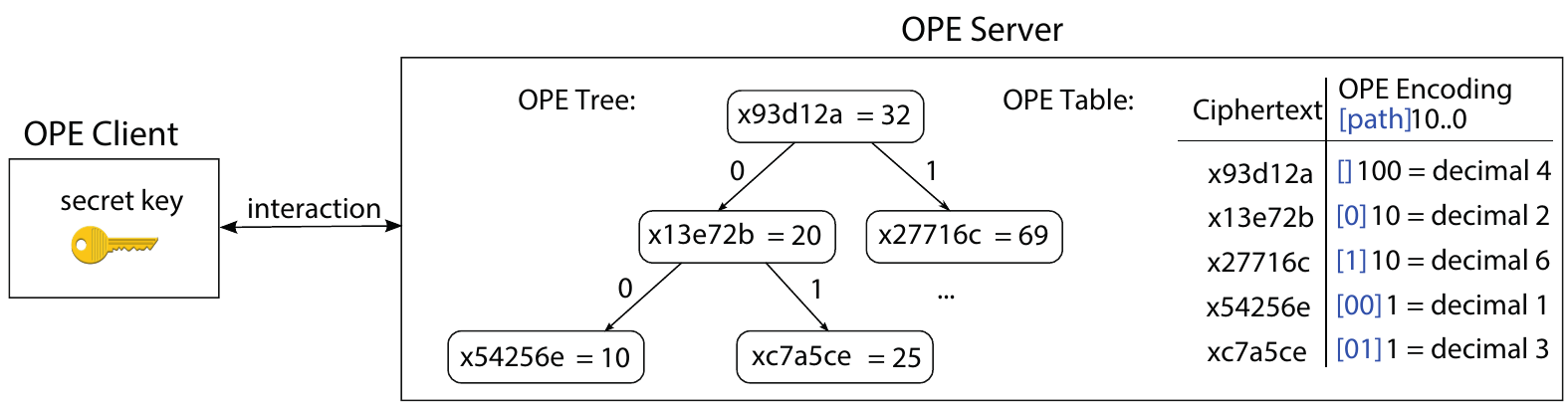}
\par\end{centering}
	\caption{Datastructure overview of mOPE \cite{compare}\label{fig:server}}
\end{figure*}

Different OPE algorithms with specific characteristics are known 
\cite{Agrawal:2004:OPE:1007568.1007632, Boldyreva:2009:OSE:1533674.1533691, Boldyreva:2011:OER:2033036.2033080, compare}. 
Mutable Order-Preserving Encoding (mOPE), a special form of OPE, is introduces by Ada 
Popa et. al. in  \cite{compare}. With mOPE, the data 
on the server is encoded in a binary search tree (see fig. \ref{fig:server}). 
The server provides the encrypted data stored in a binary 
tree and the OPE encoding path. The plaintext shown behind the encrypted
cipher text (see fig. \ref{fig:server}) is not known to the server and is provided with the purpose of the readers' information.

For our implementation we use AES, an existing deterministic encryption algorithm 
with a constant initialization vector. 
The encrypted data is represented in a binary tree, in such a way that the right 
child node is growing and the left one is decreasing. For all commercial IaaS NoSQL 
providers, it is not possible to implement own methods and algorithms for 
storing data. mOPE can be modified and used in a IaaS NoSQL scenario without 
losing security or ordering qualities. As proven in \cite{compare} the only 
information revealed of the encrypted data is the order and hence the needed minimum 
of additional exposure. Due to the high security and simplicity this 
algorithm has been chosen for the following application scenario.
The objective of this work is to propose a way to use the secure OPE algorithm mOPE 
on a cloud platform. The adaption of the mOPE algorithm is not possible without 
changing the way, data will be encrypted and sent to the cloud database due to the 
fact that cloud databases works in a different manner. In this work we discuss 
different challenges and possible solutions processing Big Data in a fast way by 
many clients. 

\section{Using Order-Preserving Encryption on Big Data}

There are different reasons for storing big amounts of data on potential untrusted 
databaseses, such as autonomous cars, intelligent homes or smart grids. However, Big 
Data is not restricted to new, expensive applications of big corporations, but 
Big Data can be acquired, also from small companies \cite{book-bigquery}.
Big Data, being generally unstructured and heterogeneous, is extremely complex to 
deal with via traditional approaches, and requires real-time or almost real-time 
analysis. \cite{7067026}
In our scenario, Big Data needs to be accessed from many 
applications or users and be processed in a fast way. This is a widely used scenario 
for any kind of data. To work with Big Data also includes the use 
of NoSQL Databases, because of the major 
advantages in processing a high amount of data in a fast way. A NoSQL 
database system is a database without a relational data model.
NoSQL is perfectly suited for processing Big Data, because of the horizontal 
scaling. Contrary to vertical scaling (e.g. SQL-Server), where a single note 
needs to be upgraded to get more computation power, horizontal scaling simply 
needs more nodes. \cite{cattell-sql} This model is cost efficient and can be 
combined with an IaaS model. Thus, extremely large data can be processed in a 
fast way. Due to the use of programming models like MapReduce 
\cite{mapreduce}, which is used in several NoSQL Databases, the processing of Big 
Data is feasible possible. With OPE on a scalable cost efficient NoSQL 
resource, no expensive infrastructure is needed and the external resource is capable 
of a secure analysis of Big Data.

\section{The Infrastructure of Mutable Order-Preserving Encryption on \\Google BigQuery}
\begin{figure*}
\noindent \begin{centering}
\includegraphics[width=0.95\textwidth]{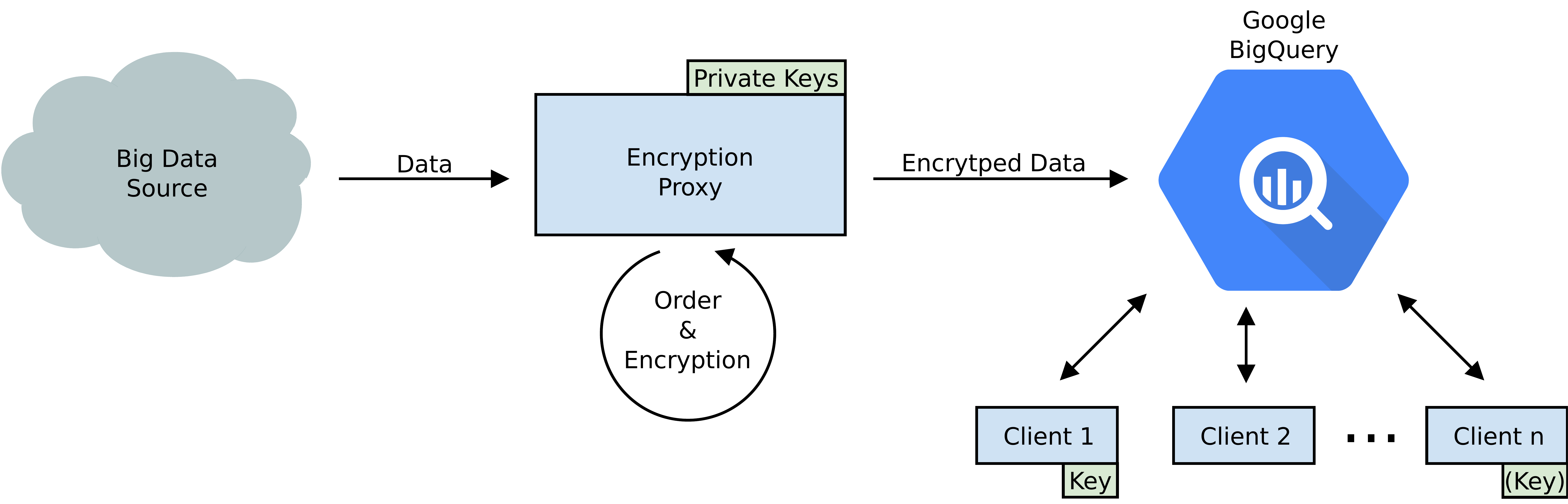}
\par\end{centering}

\caption{Concept of Order-Preserving Encrypiton on Google BigData as example IaaS provider\label{fig:concept}}
\end{figure*}

To order and encrypt data the plaintext has to be known to the system encrypting it. 
Because this work can not be outsourced to an untrusted service, we use a central 
proxy configuration. Figure \ref{fig:concept} despicts our concept for order
preserving encryption in an IaaS infrastructure. The concept of encryption includes 
an encryption proxy, an NoSQL-like IaaS and $n$ clients. The encryption proxy gets a 
Big Data chunk as input. The
proxy orders and encrypts the data by using private keys. In this concept, it 
is also possible to use different encryption algorithms and keys for 
different columns of the data. The encryption proxy in between the Big Data 
and the IaaS provider is necessary, so no additional information besides the 
order will be revealed to any external resources. The keys, used to 
encrypt and decrypt the data, will be kept secret. The sorted and encrypted 
data will then be sent to the external service. Once the  data is uploaded, 
many clients can perform analyses on it. In general, there 
is nearly no limitation on the clients or the complexity of the queries, 
because of the horizontal scaling of NoSQL Databases.
The clients themselves can have different privileges. In the example
configuration in figure \ref{fig:concept}, the client can have access
to the secret keys. This is not necessary to perform queries on the
data. If they do have the secret keys of the encryption proxy, the
full range of analysis is possible. It may be possible that encrypted
and unencrypted data are present in a single dataset. A client who does not possess 
the secret keys, can still perform analysis on unencrypted data. This
scenario is conceivable, if some of the data in a dataset is public
or available unencrypted. The unencrypted data
can then complement the encrypted data by providing additional information
and might be useful to perform other analyses. This concept allows several 
other security models and it is also possible to extend the concept for using 
different keys for different clients.
With this concept, we have created a way to use the powerful computation power of the 
cloud-service Google BigQuery by many clients but not revealing information about 
the data, which will be stored in the database. Furthermore, due to the use of 
different keys, a highly granular rights management for the clients is also possible 
which can access different information without using or implementing any rights-
management on the IaaS provider side.

\subsection{Limitations on Order-Preserving Encryption for Big Data}
Big Data usually implies that the amount of data cannot be processed by a 
single machine as the proxy in the described concept (see 
fig. \ref{fig:concept}). Also, the Big Data chunk will usually grow over 
time. This limitation does not only apply in this scenario, but is 
present every time, the order of encrypted data is needed. If new data will 
be added, the already encrypted and sorted data needs to be read from the 
external resource, decrypted, and sorted again. Outsourcing this operation to 
the external IaaS Provider does also reveal the private key to this provider. 
For security reasons, this approach is not possible. The concept we use offers the 
possibility to access data very fast because of the high power of the IaaS provider 
but still needs high computational power or time for encrypting the data on the 
encryption proxy.

\subsection{Solving the Limitations}
To solve the limitation on ordering and encrypting Big Data, we propose different 
approaches:

\begin{itemize}
\item \textbf{Using a high sorting range}: A resorting and reencryption is 
just needed if the plain text space size (domain) reaches a collision in the 
mapped sorting items (range). Then a new item cannot be sorted to the right 
position because of creating a collision. This can be prevented by choosing a 
much higher range-space. A resorting and reencryption is more unlikely. 
\item \textbf{Partial encryption}: In many cases, Big Data consists
of a mixture of public and private data. Only when private data and
public data are brought into relationship, the public data is interesting
for a potential attacker. A distinction has to be drawn between data
that should be kept secret and data, which is public available. This
can lead to a major decrease of the encrypted data by simply using an
intelligent encryption schema on the Big Data.
\item \textbf{Use of diverse encryption algorithms}: One data chunk consists
of various kinds of data, with different claims on encryption. In order to know
the right algorithm, the plaintext data needs to be analysed. If, for example,
sorting is irrelevant because a data chunk contains the name of patients,
this data can be encrypted with an probabilistic algorithm. This procedure can
be parallelised and will affect the resources of the encryption proxy less.
\item \textbf{Separating data chunks on server and merging it on client side}: We 
proposed an import of the Big Data as chunks. Assuming the returned queries 
results are significant smaller than the particular data chunk, the uploaded data 
chunks can be uploaded in different tables. A query then will be applied to the 
different chunks and the client will merge the much smaller results locally. By using 
this approach it is not necessary to reencrypt already uploaded chunks. If the 
fragmentation grows over time, a garbage collection can merge data chunks, reencrypt 
those and upload it again to the service to lower the tables on server side.\footnote
{This solution is only adequate if Big Data chunks are uploaded and the 
fragmentation is not too high.}
\end{itemize}

Additional approaches for solving the limitation can be the use of local 
NoSQL structures for the encryption and sorting proxy or the decreasing of 
the amount of data by just using the needed columns for OPE. \cite{timoschindler2016}

\section{Google BigQuery as Example IaaS Provider}
Google BigQuery is a system, designed to perform SQL-Like statements
over Big Data. Google BigQuery can query $1\frac{TB}{s}$ \cite{book-bigquery} and 
returns the result of the particular SQL statement. To achieve this, BigQuery uses 
different techniques of storing and processing data.
They are based on several main systems, like BigTable, a forerunner
of the NoSQL Database used at Google, or Megastore,
a geographically replicated, consistent NoSQL-type datastore. Megastore
uses the Paxos algorithm to ensure consistent reads and writes. \cite{book-bigquery}

The cloud service of Google utilises the Dremel Engine, a distributed
SQL query engine, to perform complex queries over Big Data. \cite{dremel}
This engine uses two technologies to achieve the goal of $1\frac{TB}{s}$:
Colossus, a large, parallel, distributed file system and ColumnIO,
a storage format which arranges the data in a manner that makes it
easier to query over this data. \cite{book-bigquery}

The Dremel Engine makes use of the Dremel Serving Tree algorithm to run the query 
on a distributed system. \cite{dremel} The possible parallelisation of SQL-
like queries is depending on the complexity of these queries. In most 
operations, the SQL query will be distributed to many workers (shards) and 
mixers. These different nodes will process the SQL query in parallel and are 
returning the result.

\section{Encrypted BigQuery Command Line Interface}
Google BigQuery offers a powerful asynchronous API to access the resources provided 
by Google. In addition, Google has released a beta client written in Python which is 
using encryption based on \cite{Popa:2011:CPC:2043556.2043566} called "Encrypted Big Query Command-Line 
Tool" (ebq). \cite{github-ebq}

To provide data for the CLI, the plaintext table data needs to be
stored either as comma-separated values (CSV) or JavaScript Object
Notation (JSON) file. The definition, which encryption should be
used on which data, needs to be provided in a second scheme file. This
extended table schema is provided by the user, where the extended
schema is a modification of the BigQuery scheme. 
The unmodified ebq-client supports Paillier's homomorphic encryption algorithm 
\cite{paillier-homomorphic}, as well as probabilistic, pseudonym and different 
searchword  algorithms. \cite{timoschindler2016}
To support Order-Preserving encryption, the client has been modified and a 
new encryption scheme is implemented. The CLI is used as encryption proxy and 
is sorting the data, using "timsort", a hybrid stable sorting algorithm, with 
a best case performance of $O\left(n\right)$, average and worst case 
performance of $O\left(n\,\log\,n\right)$. The plaintext will be encrypted 
using AES with CBC/PKCS5 Padding. The client uses a simple implementation as 
proof of concept. The encrypted data and the decimal order value will written 
together to a column separated by an special character. In addition to for 
research purpose, the order is also written to a separate column.

All encrypted data is stored in a temporary file. If the
data-encryption and -sorting is finished, the file will be uploaded by
using the API of Google BigQuery. The private key will not be transmitted
to the external source. BigQuery uses a RESTful API and the Transport-Layer-
Security (TLS) protocol to communicate encrypted with the client. The 
packages sent to the API are lossless compressed with gzip using the deflate 
algorithm. By using the compression, the data transfer can be sped
up by +50-75\%, depending on the size and the quality of the data.

We have implemented Order-Preserving Encryption based on the mOPE Algorithm.
Based on the modification, the ebq client is now capable of using a high range 
of different encryption algorithms including OPE for Big Data.

The in-depth analysis of the modified beta client in \cite{timoschindler2016} has been
investigated on different components of the client to detect possible impediments of 
the encryption. We have reviewed the resource and time consumption of uploading, 
query processing, sorting and encrypting. The analysis has proven that the bottle 
neck for the encrypted BigQuery client is the encryption of new data before 
uploading it to the external service. Due to the compression, the upload time for new 
data is relatively low, compared to the time consumption of the encryption. For 
sampling we used simulated credit card information with a sample size from $10^{3}$ 
to $10^{7}$  samples. Comparing the time consumption for ordering and encrypting the 
samples on the encryption proxy to uploading and preparing the data on Google 
BigQuery, has shown that the API runtime period is about $100$ times faster than the 
proxy. These samples also have been  tested with complex SELECT and ORDER BY 
statements. The response time for queries was in any test case less then 10 seconds\footnote{The higher time consumption is caused by the asynchronous API of Google 
BigQuery. The client will send an asynchronous call to the GBQ API. Depending on the 
load of the API at the moment, the calls will be processed one by one.}, even when 
returning $10^{6}$ query results. 

Considering Big Data, we have also evaluated that standard state of the art client 
laptop\footnote{Intel Core i7-5600 4x2.60GHz; 16 GB RAM; Arch Linux (Kernel: 4.6.2); 
Python 2.7.11} can work with encrypted Big Data. We focused on the encryption as 
bottle neck for resource consumption. For this examination we have extracted the 
encryption method used in our modified ebq client and sampled it with random 16 digit 
integer values, simulating credit card numbers. Table \ref{tab:encryption-validation} 
shows a linear correlation in the equivalent file size of the unencrypted data in 
order to the sample size as well as a linear correlation in time consumption for the 
encryption in order to the sample size. The sample scenario confirmed our expectation 
of extremely fast responses once the data is uploaded to the external system but also 
shows, that basic workstations can work with an extreme high amount of data in a 
reasonable time.

\begin{table}[h]
\renewcommand{\arraystretch}{1.3}
\caption{Sample sizes for the encryption validation\label{tab:encryption-validation}}
\noindent \begin{centering}
\begin{tabular}{lllll}
\hline 
Sample Size & Digits & Equivalent File Size & Time Consumption\tabularnewline
\hline 
$1.000$ & $16$ & $25.9$ kB & $0.0034$ s \tabularnewline
$10.000$ & $16$ & $268.9$ kB & $0.0214$ s \tabularnewline
$100.000$ & $16$ & $2.8$ MB & $0.2035$ s \tabularnewline
$1.000.000$ & $16$ & $28.9$ MB & $2.0432$ s \tabularnewline
$10.000.000$ & $16$ & $298.9$ MB & $21.7537$ s \tabularnewline
\hline 
\end{tabular}
\par\end{centering}

\end{table}

\section{Conclusion, Future Work and Discussion}
We demonstrated that it is possible to use external resources without 
decreasing the level of security more than the theoretically necessary 
minimum of Order-Preserving Encryption. With the imperative requirement of 
revealing the order of the encrypted data, it is possible to work with the 
data on the external resource, but to maintain the secrecy of private data. With the 
introduction of a new concept, which depicts to work with encryption on an external 
resource and by using the modified encrypted BigQuery client, it is feasible to 
encrypt andsort Big Data using any symmetric encryption algorithm. We have also 
combined widely used algorithms and state of the art encryption to work with modern 
Infrastructure-as-a-Service environment. OPE is still a new technology and should be 
used with care but offers already a feasible way to protect confidentiality. Further 
work will focus on overcoming the remaining limitations, using the client as 
encryption proxy. We plan to implement recent algorithm to hide the user's query 
distribution and by that making it even harder to reveal information besides the 
order.

With our solution it is possible to handle modern Big Data use cases. Even if, in 
first place, data is not considered as Big. If a system collecting timestamped GPS 
data produces a million records a day, it might not be Big Data. In three years, 
however, the system will have created a billion records. Data processing over a long 
period of time is important to detect the development or periodic trends. Often, the 
amount or the length back in time is important and will produce better results the 
more data sets are available. \cite{book-bigquery} This example shows, that a slower 
encryption but a fast analysis on the data by many clients is feasible and convenient 
for modern databases.

We have proposed different encryption algorithms which can be used with our concept 
besides Order-Preserving Encryption. Different data in a dataset does need different 
encryption algorithms depending on the projected queries and operations. This does 
have an impact on the design of the database scheme and can increases data security. 
With the modified ebq-client, the full range of algorithms and therefore operations 
is possible.

Recent work by Mavroforakis et. al. \cite{Mavroforakis:2015:MOE:2723372.2749455} has 
revisited the Modular Order-Preserving Encryption (MOPE) Algorithm \cite
{Boldyreva:2011:OER:2033036.2033080}. MOPE is slightly different to mOPE, but is is 
also possible for a external database service to gain more information besides the 
order by observing the user's queries. In \cite
{Mavroforakis:2015:MOE:2723372.2749455} the author discusses three contributions to 
hide the user's query distribution by mixing it with another distribution. The 
described methods are possible, but are not implemented yet in the proof of concept 
client of this work.

For high security needs, it is also possible to use Oblivious RAM (ORAM). ORAM is a 
technique that hides all information about which positions in an outsourced database 
are accessed by the client, by continually shuffling around and re-encrypting the 
data. \cite{Stefanov:2013:POE:2508859.2516660, Goldreich:1996:SPS:233551.233553, 
Mavroforakis:2015:MOE:2723372.2749455}. Considering \cite
{Mavroforakis:2015:MOE:2723372.2749455} ORAM is less efficient and the proposed 
solutions more sufficient for OPE.

\bibliographystyle{bib_style/unsrtdin.bst}
\bibliography{bibliography}

\end{document}